\documentclass[submission,copyright,creativecommons]{eptcs}
\providecommand{\event}{Formal Methods for Autonomous Systems 2026}

\usepackage{paralist}
\usepackage[T1]{fontenc}
\usepackage[utf8]{inputenc}
\usepackage{underscore}
\usepackage{amsmath,amssymb}

\usepackage[dvipsnames]{xcolor}
\usepackage[color]{zed}
\newcommand{\clz}{\color{ZedColor}}
\newcommand{\clmb}{\color{MidnightBlue}}
\newcommand{\codex}[1]{\texttt{#1}}

\usepackage{booktabs,tabularx,array}
\usepackage{listings}
\usepackage{tikz}
\usetikzlibrary{arrows.meta,positioning,fit,calc}
\usepackage{microtype}
\usepackage{placeins}
\usepackage{float}
\usepackage{url}
\hypersetup{
    colorlinks=true,
    linkcolor=red,
    citecolor=blue,
    urlcolor=cyan,
    pdftitle={Contract-Aware Rescue of a Drifted Isabelle Development: The Double-Tank Case Study},
    pdfauthor={Jim Woodcock; Gabriel Leite; Augusto Sampaio; Ran Wei},
    pdfkeywords={Isabelle/HOL, large language models, proof repair, theory drift, hybrid systems, CAPRI}
}

\newcommand{\code}[1]{\nolinkurl{#1}}

\lstdefinelanguage{YAML}{
  keywords={true,false,null},
  sensitive=false,
  comment=[l]{\#},
  morestring=[b]"
}
\lstdefinelanguage{Isabelle}{
  keywords={theorem,lemma,definition,assumes,shows,and,where,if,then,else,proof,qed,by},
  sensitive=true,
  morecomment=[s]{(*}{*)},
  morestring=[b]"
}
\title{Contract-Aware Rescue of a Drifted Isabelle Development: \\ The Double-Tank Case Study}

\author{
  Jim Woodcock
  \institute{Southwest University, China; Aarhus University, Denmark; University of York, UK}
  \email{jim.woodcock@york.ac.uk}
  \and
  Gabriel Leite
  \institute{Universidade Federal de Pernambuco, Brazil}
  \email{gnl2@cin.ufpe.br}
  \and
  Augusto Sampaio
  \institute{Universidade Federal de Pernambuco, Brazil}
  \email{acas@cin.ufpe.br}
  \and
  Ran Wei
  \institute{Lancaster University, UK}
  \email{r.wei@lancaster.ac.uk}
}
\def\titlerunning{Contract-Aware Rescue of a Drifted Isabelle Development}
\def\authorrunning{J. Woodcock et al.}
\date{}

\begin{document}

\maketitle

\begin{abstract}

  Large language models can propose proofs for interactive theorem provers, but a successful build does not show the surrounding verification task was preserved. We study this problem in an Isabelle development of a sampled-data double-tank controller. The work began with nine theories and ten unfinished obligations, grew to a 16-theory build without \texttt{\clmb sorry}, \texttt{\clmb oops}, added axiomatisation, or oracle use, and accumulated 23 stable and 36 broken proof states. A retrospective audit found material changes in 16 of the 100 original declarations, including a weakened end-to-end assurance theorem that assumed three of the four requirements in its conclusion. We used CAPRI, a contract-aware proof-repair tool, to govern a reconstruction by combining Isabelle acceptance with an independent check of repository changes against machine-readable edit contracts. The reconstruction discharged all ten scoped obligations within the original nine-theory structure. A secondary replay by a co-author reproduced the R10 build, contract checks, control tests, and principal audit findings; independent replication remains future work. Operational end-to-end verification remains incomplete: we still need to connect operational executions to the reconstructed quantitative trace contract, a task requiring an extended contract.

\end{abstract}


\section{Introduction}

In theorem proving, a result is accepted only if its proof can be checked from the definitions, assumptions, and previously proven results available to the prover~\cite{PaulsonNW2019}. LLMs can assist by generating proof scripts, proposing useful lemmas, and revising suggestions in response to prover diagnostics~\cite{JiangLTCOMWJ2022,FirstRRB2023,PoluS2020,YangSGCSYGPA2023}. Proof checking remains with the theorem prover; the LLM supplies candidate text. Trouble begins when the model, or an iterative repair tool built around it, can also edit material outside the intended proof body. A repository can compile after weakening a theorem, adding a convenient assumption, changing a definition or import, or editing an unrelated file. The formal problem may then differ from the one assigned. Isabelle still checks the submitted theory correctly, through the LCF-style separation between proof-producing automation and a small trusted inference kernel~\cite{PaulsonNW2019,BohmeN2010}; the kernel does not account for which parts of the repository the developer authorised the repair process to change.

The double-tank development exposes this distinction in a concrete setting. It is based on an earlier formal design of a double-water-tank controller~\cite{ThaeH2001}: two tanks are connected in cascade, a pump controls the inflow, liquid levels evolve continuously, and a discrete controller acts at periodic samples. The Isabelle/HOL theories cover the plant, controller, temporal requirements, approximate conformance, runtime monitoring, and institution-based semantics. We began with nine files containing the vocabulary and principal theorem statements, leaving the difficult proofs open. We used ChatGPT to explore proof strategies. It found some successful proofs, alongside many failed attempts involving continuous evolution, invariant preservation, relational framing, and tolerance transfer. 

The repository eventually contained 16 theories and built without using \texttt{\clmb sorry}, \texttt{\clmb oops}, added axiomatisation, or oracles (collectively, \emph{admissions}). Later, a comparison with the original outline revealed changes to definitions, assumptions, boundary conventions, and public claims. Isabelle accepted the resulting theory, but it no longer matched the initial task in every respect. Three terms distinguish the relevant boundaries. The \emph{theory frame} is the set of statements, definitions, assumptions, imports, file structure, and public claims that delimit the authorised proof task. \emph{Theory drift} is an accumulation of individually accepted edits that changes that task without an explicit change of intent. A \emph{semantic seam} is a boundary between separately formalised layers that needs an explicit bridge before their meanings can be treated as connected. Drift may weaken an engineering claim or make an unresolved seam hard to see.

The paper records the proof-discovery history of the hybrid systems development, including 23 stable proof states and 36 distinct broken states. The failed candidates influenced later decisions about the model, its assumptions, and the supporting lemmas, so they form part of the evidence rather than discarded intermediate work. A declaration-level audit measures the distance between the completed sixteen-theory repository and the first double-tank outline. Among 100 named declarations, 84 preserved identical formal content. The remaining 16 differed in declaration kind, definition, assumptions, domain, or boundary treatment.

The reconstruction kept the first theory frame while drawing on mathematical results from the larger repository. Where the semantic chain was incomplete, we named the missing link and assigned it a successor obligation instead of moving it into the assumptions of the theorem under repair. CAPRI checks the declared authority boundary, whereas semantic equivalence or modelling adequacy remains a developer judgement requiring an explicit successor contract when the task changes. The study follows one evolving double-tank repository and makes no comparison between alternative CAPRI repair configurations.

We review related work, the double-tank system, its Isabelle development, and the archived proof-discovery process. Later sections introduce CAPRI, report the drift audit and controlled reconstruction, and state the remaining operational assurance obligation before discussing the study's limitations.

\section{Related work}

Isabelle follows the LCF tradition: automation may search widely, but accepted theorems ultimately depend on kernel-checked inferences~\cite{PaulsonNW2019}. Sledgehammer reconstructs externally found arguments in Isabelle~\cite{BohmeN2010}; learning-based systems such as Thor, Baldur, GPT-f, and LeanDojo extend proof search or repair with language models~\cite{JiangLTCOMWJ2022,FirstRRB2023,PoluS2020,YangSGCSYGPA2023}. CAPRI addresses a different question. Given a candidate repository, it checks whether changes are confined to an authorised edit boundary. This complements proof transport and maintenance techniques such as Pumpkin Pi and large-scale Isabelle proof engineering~\cite{RingerPYLG2021,RingerPSGT2019,Andronick2019}: CAPRI does not establish semantic equivalence of an authorised change, but prevents unrecorded changes to protected statements, definitions, assumptions, imports, and files.

Theory drift connects to requirements evolution and traceability~\cite{NuseibehE2000,GotelF1994}. Theory change is normal; drift occurs when accepted edits change the engineering question without an explicit decision or replacement obligation. Our correspondence map records what changed; CAPRI records which changes were permitted.

Hybrid-systems assurance provides the application setting, while existing calculi supply the semantic foundations. Differential dynamic logic, Isabelle-based hybrid predicate-transformers, Isabelle/UTP, and Hybrid Relations provide relevant foundations~\cite{QueselMLAP2016,MuniveS2019,FosterZW2014,foster2019hybrid}. Institution theory, duration calculus, runtime verification, ModelPlex, and trace-conformance metrics supply related abstractions for semantic transport, duration, execution monitoring, model-to-execution validation, and bounded implementation deviation~\cite{GoguenB1992,ChaochenH2004,LeuckerS2009,MitschP2014,DeshmukhMP2017}. We study how such semantic choices, requirements, and bridges are protected during iterative proof work.

The acceptance decomposition can transfer to other provers: their native checker supplies \codex{\clmb Build}, while a repository-side policy supplies \codex{\clmb Conforms}. A Lean or Coq port would need prover-specific adapters for declaration and proof-region location and for dependency/build metadata. The principle is prover-agnostic; the implementation and evaluation here are Isabelle-specific.

\section{The double-tank case study}


The case study builds on the double-water-tank of Hong Ki Thae and Dang Van Hung~\cite{ThaeH2001}. A pump feeds the upper tank; water passes into the lower tank and is then discharged. At each sample, the controller chooses the pump command used during the next continuous phase to track the lower-tank reference while keeping both levels within their limits. The policy is fixed rather than adaptive. Even this compact plant brings together continuous dynamics, discrete control, sampling, timing, physical bounds, disturbance and sensor tolerances, and the gap between a mathematical model and its implementation.

Between samples, the tank levels follow the plant differential equations. A sampling action records observations; the controller then applies its fixed policy, updates the pump command, and begins the next continuous phase. The state includes the pump signal, controller mode, sampled values, target reference, event information, timing, and the two levels. Lens-based state access from Isabelle/UTP~\cite{FosterZW2014} supports local updates and makes frame conditions explicit. Those frame conditions are essential to the closed-loop invariant: each proof must account for state components changed by an action and those preserved by it.

\subsection{Requirements and assurance argument}

The development has four requirements:
\begin{inparaenum}[(1)]
\item \textbf{No-overflow safety}: both levels remain within their physical bounds;
\item \textbf{Bounded settling}: the system enters a steady tracking region within a bounded delay after a reference event;
\item \textbf{Stable tracking}: that region persists while the reference stays unchanged;
\item \textbf{Availability}: acceptable behaviour for a sufficient measured duration over an observation horizon.
\end{inparaenum}
Together, these mix invariant safety, bounded response, persistence, and quantitative duration, so they exercise different semantic mechanisms and expose different risks of accidental weakening.

The safety argument follows the control cycle. Sampling preserves physical admissibility. The controller leaves the physical state unchanged and establishes the guard for the next plant phase. Under the stated flow and compatibility assumptions, plant evolution preserves non-negativity and capacity. These component results yield preservation for one complete cycle, and finite induction gives the corresponding invariant for repeated cycles. A trace bridge carries reachable-state safety to the trace requirement. The broader development builds on this invariant proof with temporal interpretation, institution-level representation, approximate conformance, and executable monitoring~\cite{GoguenB1992,ChaochenH2004,LeuckerS2009,MitschP2014,DeshmukhMP2017}.

The case study directly supports automated system assurance because it separates the physical plant, fixed controller, verified guards, runtime monitors, and formal evidence, a structure that recurs in automated fuel transfer, coolant circulation, battery thermal management, hydraulic regulation, and ballast or trim control. In each case, the physical equations may be straightforward, yet assurance still needs a precise account of assumptions, operating envelope, controller authority, monitoring, and evidence provenance. The architecture could support later semi-autonomous or autonomous extensions, but this contribution focuses on the automated control system itself.

\subsection{Isabelle architecture and development baselines}

The first skeleton contained nine theories. \codex{\clmb Double_Tank_Alphabet} introduced the state, traces, relations, and lens structure. \codex{\clmb Double_Tank_Plant}, \codex{\clmb Double_Tank_Sampling}, and \codex{\clmb Double_Tank_Controller} represented the main operational components. \codex{\clmb Double_Tank_Requirements} defined safety, settling, tracking, availability, and temporal properties. \codex{\clmb Double_Tank_Closed_Loop} composed the relations and lifted one-cycle preservation using finite iteration. \codex{\clmb Double_Tank_Conformance} introduced distance, envelopes, and tolerance transfer, while \codex{\clmb Double_Tank_Monitor} connected the denotational requirements to executable checks. The entry theory \codex{\clmb Double_Tank} exported public assurance results.

Three semantic seams were visible in the first outline: \codex{\clmb solves_tank_ode} stood in for continuous ODE or Hybrid UTP semantics, \codex{\clmb state_duration} for duration-calculus or interval-measure semantics, and \codex{\clmb is_closed_loop_trace} for the relationship between real-time traces and repeated plant, sampling, and controller execution. Leaving these seams explicit avoided importing a second semantic foundation merely to populate the outline. It also created a risk: when a seam was later defined, its definition had to capture the intended semantics rather than make the target theorem true by construction.

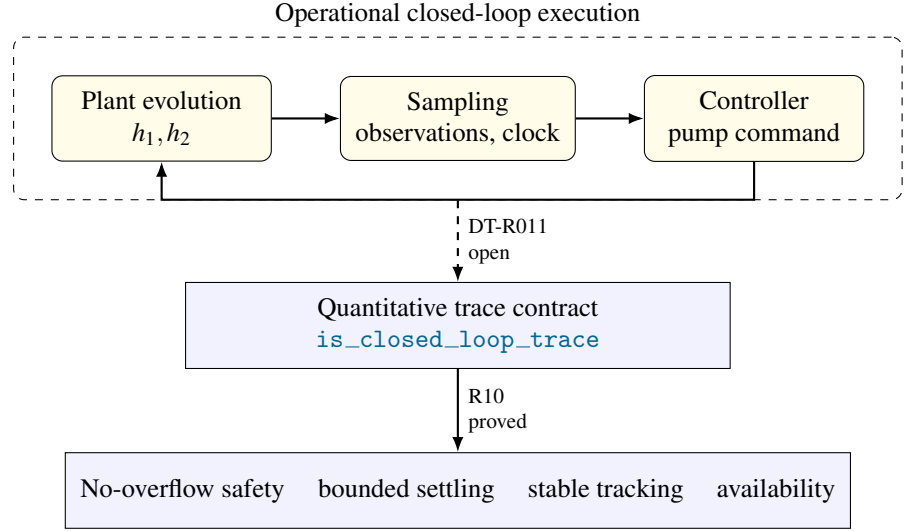
\begin{figure}[!h]
  \centering
  \begin{tikzpicture}[
    component/.style={draw, rounded corners, align=center, minimum width=29mm, minimum height=9mm, inner sep=2mm, font=\small, fill=yellow!10},
    semantic/.style={draw, align=center, minimum width=72mm, minimum height=10mm, inner sep=2mm, font=\small, fill=blue!5},
    arrow/.style={-{Latex[length=2mm]}, thick},
    openarrow/.style={-{Latex[length=2mm]}, thick, dashed}
  ]
    \node[component] (plant) {Plant evolution\\$h_1,h_2$};
    \node[component, right=9mm of plant] (sampling) {Sampling\\observations, clock};
    \node[component, right=9mm of sampling] (controller) {Controller\\pump command};
    \draw[arrow] (plant) -- (sampling);
    \draw[arrow] (sampling) -- (controller);
    \draw[arrow] (controller.south) |- ++(0,-5mm) -| (plant.south);
    \node[draw, dashed, rounded corners, fit=(plant)(sampling)(controller), inner sep=5mm,
          label={[font=\small]above:Operational closed-loop execution}] (cycle) {};

    \node[semantic, below=16mm of sampling] (contract) {Quantitative trace contract\\\texttt{\clmb is\_closed\_loop\_trace}};
    \node[semantic, below=11mm of contract] (requirements) {No-overflow safety \quad bounded settling \quad stable tracking \quad availability};

    \draw[openarrow] (cycle.south) -- node[right, align=left, font=\scriptsize] {DT-R011\\open} (contract.north);
    \draw[arrow] (contract.south) -- node[right, align=left, font=\scriptsize] {R10\\proved} (requirements.north);
  \end{tikzpicture}
  \caption{Operational and assurance layers, read top-to-bottom. Plant, sampling, and controller steps form repeated closed-loop executions; the open DT-R011 bridge must show that their coherent traces satisfy \codex{\clmb is_closed_loop_trace}; R10 then derives the four requirements from that quantitative trace contract.}
  \label{fig:operational-assurance}
\end{figure}

In the final reconstruction, milestone R10 establishes the four requirements from the quantitative trace contract; the remaining operational-to-trace bridge is recorded separately as DT-R011. Figure~\ref{fig:operational-assurance} makes this remaining assurance boundary explicit. The operational theories define repeated plant, sampling, and controller execution. The reconstructed trace contract provides the semantic interface used by the verified reconstruction. What remains to be established is that coherent traces generated by the operational model satisfy this interface.

Table~\ref{tab:chronology} distinguishes the principal baselines used in the project. The labels refer to different artefacts and should not be confused. The v0.24 baseline preserves and audits the expanded sixteen-theory development, whereas R1--R10 is a separate reconstruction within the first nine-theory architecture.

\begin{table}[!h]
  \centering
  \caption{Development chronology. The v0.x labels denote historical development baselines; R1--R10 denote the CAPRI reconstruction milestones detailed in Table~\ref{tab:milestones}.}
  \label{tab:chronology}
  \small
  \renewcommand{\arraystretch}{1.5}
  \begin{tabularx}{\textwidth}{
    @{}
    >{\raggedright\arraybackslash}p{0.20\textwidth}
    X
    @{}
    }
    \toprule
    Stage                    & Characterisation \\
    \midrule
    Original outline / v0.1  & Nine-theory skeleton containing the intended vocabulary and ten unfinished obligations. \\
    v0.10                    & Nine theories, 953 lines, 40 definitions, 28 lemmas, 19 theorems, ten \codex{\clmb oops}, and no \codex{\clmb sorry}. \\
    v0.23                    & Expanded sixteen-theory development of about 6,250 lines, accepted by Isabelle and admission-free; no-overflow safety was proved, while settling, stable tracking, and availability remained conditional or unresolved. \\
    v0.24 CAPRI baseline     & The v0.23 mathematics together with frozen provenance, correspondence data, contracts, architectural invariants, and a drift audit. \\
    Reconstruction R1--R10   & A separate CAPRI-governed reconstruction within the first nine-theory architecture, using contracts derived from the earliest theory frame. \\
    R10 final reconstruction & All ten scoped reconstruction obligations discharged; Isabelle and contract checks accepted; the final requirement theorem remains relative to \codex{\clmb is_closed_loop_trace}. \\
    DT-R011                  & Planned operational trace-adequacy bridge. \\
    \bottomrule
  \end{tabularx}
\end{table}

The expanded v0.23 bundle added the theories \codex{\clmb Duration_Semantics}, \codex{\clmb Double_Tank_Duration}, \codex{\clmb Hybrid_Institution}, \codex{\clmb Double_Tank_Institution}, \codex{\clmb Thermostat_Institution}, \codex{\clmb Hybrid_Trace_Institution}, and \codex{\clmb Hybrid_Temporal_Institution}, drawing on Duration Calculus and institution theory respectively~\cite{GoguenB1992,ChaochenH2004}. Its sixteen theories fell into four layers: generic institution and duration foundations; the concrete double-tank model; institution-parametrised safety and temporal representations; and the conformance, monitoring, and entry interfaces. To check that the generic institution machinery had not been inadvertently specialised to the double-tank case, we also instantiated it with an independent thermostat model. The successful thermostat instance therefore serves as a generality check on the generic institution machinery. The supplementary artefact records the complete theory catalogue.

\section{LLM-guided proof discovery}

The development was theory-first: the nine initial theories fixed the vocabulary and headline statements before attempting the difficult proofs. They contained ten unfinished obligations. The intended proof order followed semantic dependency: continuous evolution and plant invariants first, then sampling and control, finite closed-loop iteration, conformance, temporal requirements, monitors, and finally the end-to-end theorem. This ordering later became the basis of the controlled reconstruction.

ChatGPT was used in a diagnostic loop similar to prover-checked generation workflows~\cite{JiangLTCOMWJ2022,FirstRRB2023}. Candidate proofs or supporting lemmas were installed in disposable repositories, and the corresponding complete Isabelle2025-2 session was built; failure diagnostics informed later attempts. Isabelle determined proof acceptance, while the developer selected subsequent baselines. At this stage of the development, the surrounding theory frame was not independently protected. The archive preserves candidate repositories and diagnostics, although it lacks a stable model identifier for every historical ChatGPT interaction.

After excluding byte-identical downloads and packaging-only changes, the archive contains 59 substantive states: 23 stable and 36 broken. Failures exposed unavailable facts, hidden analytic side conditions, weak induction hypotheses, missing frame conditions, and mismatches between the intended physical argument and the formal goal. Several obstacles lay outside the proof body (for example, an undefined semantic seam or a statement requiring reformulation), which are precisely the points at which unrestricted editing can turn repair into theory drift. By v0.23, the enlarged sixteen-theory development built without \codex{\clmb oops}, \codex{\clmb sorry}, or axiomatisations, but bounded settling, stable tracking, availability, and the full end-to-end assurance claim, was still conditional or unresolved.

\section{From proof repair to CAPRI}

CAPRI treats proof repair as a controlled repository transformation~\cite{WoodcockLSW2026}. For baseline repository $\clz R$, candidate $\clz R'$, and edit contract $\clz C$, acceptance is
\begin{zed}
  \operatorname{Accept}(R,R',C)
  \;\Longleftrightarrow\;
  \operatorname{Build}(R') \land \operatorname{Conforms}(R,R',C).
\end{zed}
Isabelle establishes $\clz Build$; an independent checker establishes $\clz Conforms$. The checker compares the frozen baseline and candidate using repository-, file-set, and source-region checks rather than semantic AST equivalence: only nominated regions may change; protected statements, assumptions, definitions, imports, files, and build metadata remain fixed as required by the contract; and forbidden tokens are rejected. The same check runs before and after the Isabelle build. Thus $\clz Conforms$ detects unauthorised structural edits but cannot judge semantic equivalence inside an authorised region; a permitted rename, scope change, or replacement definition still requires explicit wider authority and substantive semantic review.

\begin{figure}[htbp]
  \begin{minipage}[t]{0.48\textwidth}
\begin{lstlisting}[language=YAML,backgroundcolor=\color{gray!10},basicstyle=\ttfamily\scriptsize,aboveskip=0.2em,belowskip=0.2em]
target:
  theory: Double_Tank_Plant
  theorem: plant_preserves_nonnegative_levels
editable:
  proof_body: true
protected:
  theorem_statement: true
  assumptions: true
  definitions: true
  imports: true
  other_files: true
forbidden: [sorry, axiomatisation, oracle]
\end{lstlisting}
  \end{minipage}\hfill
  \begin{minipage}[t]{0.48\textwidth}
\begin{lstlisting}[language=YAML,backgroundcolor=\color{gray!10},basicstyle=\ttfamily\scriptsize,aboveskip=0.2em,belowskip=0.2em]
baseline: R9
task: R10
editable_files:
  - Double_Tank_Plant.thy
  - Double_Tank_Requirements.thy
  - Double_Tank_Closed_Loop.thy
  - Double_Tank_Conformance.thy
protected:
  all_other_formal_files: true
required_checks:
  - complete_isabelle_build
  - contract_conformance
\end{lstlisting}
  \end{minipage}
  \caption{CAPRI authority boundaries: representative proof-only contract (left) and schematic, non-verbatim R9-to-R10 wider contract (right). The latter explicitly limits edits to four formal files.}
  \label{fig:contracts}
\end{figure}

\noindent The checker distinguishes a legitimate accepted repair from safe failure and from \emph{false success}: a candidate building only after an unauthorised repository change. A control where the target theorem was weakened to a trivial statement demonstrated the distinction: Isabelle accepted the altered theory, while CAPRI rejected the delta. The double-tank project then used this build-plus-conformance criterion to reconstruct a development in which drift had accumulated through a longer sequence of locally reasonable edits.

\section{Auditing theory drift}


Isabelle continued to accept the expanded theories, so the audit focused on correspondence to the first outline rather than kernel soundness. Its universe was the 100 named declarations in the earliest frame; the 295 names added in v0.23 were recorded separately. Machine-generated inventories and correspondence outputs supplied the candidate mapping; the researchers then inspected every non-identical case and each rename, split, or merge and recorded a relationship class and rationale. \emph{Identical formal content} means that the formal statement or definition body was unchanged, excluding presentation-only edits. Eighty-four declarations met that test, and sixteen did not. The 16/100 result therefore combines mechanical comparison with manual semantic classification: the archived map and rationales are inspectable, but the classifications remain interpretive and were not independently adjudicated. Among the ten original unfinished obligations, six were proved unchanged and three only after reformulation. The original unconditional end-to-end obligation remained open.

The audit did not count every difference as a defect. An uninterpreted semantic seam eventually needs a definition, and several proofs required explicit flow guards or frame conditions. Other edits altered a requirement or public assurance claim and therefore called for correction or prominent qualification. Table~\ref{tab:drift-groups} groups the sixteen changes by their effect on the engineering argument.

\begin{table}[!h]
  \centering
  \caption{Material changes among the 100 original declarations.}
  \label{tab:drift-groups}
  \footnotesize
  \renewcommand{\arraystretch}{1.5}

  \begin{tabularx}{\textwidth}{
    @{}
    p{0.18\textwidth}
    p{0.07\textwidth}
    p{0.35\textwidth}
    X
    @{}
    }
    \toprule
    Group                  & Count & Representative declarations                                                                                 & Engineering effect \\
    \midrule
    Semantic instantiation / refactoring & 4     & \codex{\clmb solves_tank_ode}, \codex{\clmb state_duration}, \codex{\clmb availability_requirement}, \codex{\clmb is_closed_loop_trace} & Supplies missing meaning; \codex{\clmb availability_requirement} is also refactored through \codex{\clmb availability_requirement_with}. \\
    Plant and assumptions  & 4     & \codex{\clmb plant_evolution}, guarded capacity, one-cycle and finite-iteration preservation                       & Makes the preservation argument sounder while restricting the admitted plant and parameter domain. \\
    Safety domain and boundary & 6     & \codex{\clmb safe_at}, \codex{\clmb safety_requirement}, overflow monitor and its soundness/coverage results              & Changes strict $<$ to $\leq$ and restricts \codex{\clmb safety_requirement} to non-negative time. \\
    End-to-end interface   & 2     & \codex{\clmb closed_loop_meets_requirements}, \codex{\clmb double_tank_end_to_end_assurance}                              & Moves settling, stability, and availability into premises of the public assurance claim. \\
    \bottomrule
  \end{tabularx}
\end{table}

\subsection{The principal assurance failures}

The most serious drift concerned the safety boundary and the end-to-end theorem. Changing \texttt{\clmb <} to $\leq$ at tank capacity looks minor but shifts the engineering interpretation of both the requirement and the monitor. Under the prior strict policy, reaching capacity already lies outside the safe region, leaving no margin for disturbance, measurement uncertainty, or numerical error; under the revised policy, equality is accepted and the monitor reports \codex{\clmb Monitor_OK}. The revised definitions and monitor theorems were internally consistent, yet they certified a weaker policy than the first outline.

The change to \codex{\clmb closed_loop_meets_requirements} was more consequential. The first development was meant to show that the closed-loop controller established safety, bounded settling, stable tracking, and availability. In v0.23, safety was derived, but the other three requirements appeared as premises; hence the theorem was logically valid but no longer discharged the intended controller-assurance obligation. Exporting the corresponding conditional result as \codex{\clmb double_tank_end_to_end_assurance} invited a reader to infer that the full requirement set had been proved. This is a characteristic form of drift: the theorem name and its architectural position preserve the appearance of the first claim even though the logical burden has quietly moved into the assumptions.

Semantic seams create a different risk. Giving \codex{\clmb solves_tank_ode} and \codex{\clmb state_duration} concrete meanings was necessary progress, but each choice committed the development to a particular interpretation of the previously abstract seam. Defining \codex{\clmb is_closed_loop_trace} by pointwise reachability was more problematic, since a set of individually reachable observations need not form one coherent execution. A trace theorem can then become easy to prove if the trace predicate already includes the desired property or admits observations independently of their temporal relationship. The audit treated such definitions as semantic changes demanding explicit provenance and a successor obligation connecting them to operational implementation.

The two most consequential changes appear schematically below. The exact Isabelle declarations contain supplementary context, but the extracts below show the logical difference.

\begin{lstlisting}[language=Isabelle,backgroundcolor=\color{gray!10}]
(* Original boundary policy *)
safe_at p s ≡
  0 ≤ h1 s ∧ h1 s < capacity1 p ∧
  0 ≤ h2 s ∧ h2 s < capacity2 p

(* Drifted boundary policy *)
safe_at p s ≡
  0 ≤ h1 s ∧ h1 s ≤ capacity1 p ∧
  0 ≤ h2 s ∧ h2 s ≤ capacity2 p
\end{lstlisting}

\begin{lstlisting}[language=Isabelle,backgroundcolor=\color{gray!10},basicstyle=\ttfamily\footnotesize,aboveskip=0.45em,belowskip=0.45em]
(* Intended end-to-end obligation *)
assumes "admissible_parameters p"
    and "is_closed_loop_trace p tr"
shows "safety_requirement p tr
     ∧ settling_requirement p tr
     ∧ stable_tracking_requirement p tr
     ∧ availability_requirement p tr"

(* Drifted conditional result *)
assumes "admissible_parameters p"
    and "settling_requirement p tr"
    and "stable_tracking_requirement p tr"
    and "availability_requirement p tr"
shows "safety_requirement p tr
     ∧ settling_requirement p tr
     ∧ stable_tracking_requirement p tr
     ∧ availability_requirement p tr"
\end{lstlisting}

\subsection{Architectural drift}

The generic institution machinery, duration semantics, common safety and temporal languages, conformance layer, monitors, and thermostat instance all added useful mathematics. Their introduction also created new ways for the concrete double-tank claim to change. An institution layer might replace the source semantics instead of representing it; a direct theorem might bypass satisfaction; a translation might alter the meaning of a requirement; or a conditional result might be exported under an unconditional name. The v0.24 CAPRI baseline therefore fixed architectural invariants. Institution layers must preserve the source semantics, transported results must depend on a proved satisfaction condition, the four requirements must have traceable source-to-target mappings, and public all-requirements claims must expose any premises that remain unproved. The declaration map and semantic audit supply information absent from both the \codex{\clmb Build} and \codex{\clmb Conforms} verdicts: whether an authorised change still preserves the engineering meaning of the original task.

\section{Contract-governed rescue}

\subsection{Preserving intention and proof knowledge}

The rescue froze two artefacts. The original nine-theory frame retained the intended definitions, requirements, theorem statements, and semantic seams. The verified v0.23 repository retained the accumulated mathematics, reusable proofs, and failed approaches that had exposed hidden assumptions, a familiar proof-maintenance issue in evolving Isabelle developments~\cite{Andronick2019}. The v0.24 CAPRI baseline added provenance records, declaration correspondence, architectural constraints, contracts, public-interface checks, manifests, and hashes. Both sources were needed: the first supplied the task boundary, and the second supplied proof knowledge that would otherwise have been lost. 

We mapped each significant original declaration to its v0.23 counterpart and classified it as preserved, renamed, specialised, generalised, split, merged, reformulated, superseded, missing, rejected, or added. The four principal requirements received end-to-end traceability from physical motivation through Isabelle definition, supporting assumptions, theorems, institution translation, and runtime monitor, making boundary changes and conditional assurance claims visible right where they affected the public interface.

The reconstruction proceeded in a separate source tree. For a proof-only task, the contract froze the theorem statement, assumptions, locale context, definitions, imports, file set, protected text outside the proof, Isabelle version, and build command, so only the nominated proof region could change. When an obligation could not be solved honestly under that authority, the workflow did not silently widen the patch; it required a new contract for a statement change, semantic extension, architectural change, or conservative new theory, and that broader contract had to name the permitted change and retain the existing obligation in the record.

Candidates were checked against the contract before Isabelle ran, built in the frozen environment, and checked again after the build. The first check kept obviously unauthorised candidates from consuming prover time; the second caught changes introduced by patch application or an iterative tool outside the displayed proof proposal. A candidate could therefore end in one of three useful states: an accepted repair, where build and contract checks both passed; a safe failure, where the proof failed but the protected frame stayed unchanged; or a false success, where Isabelle accepted the repository but the independent checker found an unauthorised change.


The controlled reconstruction discharged ten scoped obligations in dependency order. Each accepted state became the protected baseline for the next task, so the authority to change the development expanded only through an explicit contract. Table~\ref{tab:milestones} summarises the sequence.

\begin{table}[!t]
  \centering
  \caption{CAPRI-governed reconstruction milestones.}
  \label{tab:milestones}
  \small
  \renewcommand{\arraystretch}{1.5}
  \begin{tabularx}{\textwidth}{
    @{}
    >{\centering\arraybackslash}
    p{0.07\textwidth}
    >{\raggedright\arraybackslash}
    p{0.32\textwidth}
    X
    @{}
    }
    \toprule
    State & Obligation                                        & Role in the reconstruction \\
    \midrule
    R1    & \codex{\clmb plant_zero_duration}                        & Supplied a usable plant semantics and handled the empty derivative interval at zero duration. \\
    R2    & \codex{\clmb plant_preserves_nonnegative_levels}         & Established scalar barrier reasoning preventing either level from crossing below zero. \\
    R3    & \codex{\clmb plant_preserves_capacity_under_guard}       & Derived capacity preservation from explicit controller and plant guards rather than building safety into evolution. \\
    R4    & \codex{\clmb sampling_occurs_at_period}                  & Repaired sampling-clock progress whilst preserving the sampling frame. \\
    R5    & \codex{\clmb trace_tolerance_budget}                     & Proved additive composition of trace tolerances. \\
    R6    & \codex{\clmb approximate_safety_transfer}                & Transferred strong safety from the ideal process to an approximately conforming implementation. \\
    R7    & \codex{\clmb approximate_tracking_transfer}              & Transferred tracking with an explicit accumulated tolerance budget. \\
    R8    & \codex{\clmb closed_loop_cycle_preserves_physical_state} & Connected plant framing, controller guards, sampling, and one complete cycle. \\
    R9    & \codex{\clmb closed_loop_preserves_physical_state}       & Lifted one-cycle preservation using finite iteration. \\
    R10   & \codex{\clmb closed_loop_meets_requirements}             & Completed the reconstructed composition using explicit duration semantics and a non-circular trace contract; operational adequacy remains DT-R011. \\
    \bottomrule
  \end{tabularx}
\end{table}

\noindent We set edit authority separately for each reconstruction milestone. Proof-only contracts were used when possible. Semantic completion or multi-file composition required a newly named contract that recorded the wider authority and retained the predecessor obligation. The archived contracts, rather than theorem names, determine the editable regions and protected files.

Because reconstruction followed the dependency order, the final theorem had to use the established plant, sampling, controller, conformance, and iteration results. Its statement could not be relaxed merely because a proof attempt failed. The R10 archive illustrates the point. Two candidates used unavailable facts or methods (\codex{\clmb diff_nonneg}, \codex{\clmb sub_nonneg}, and \codex{\clmb nlinarith}) and left the conformance argument incomplete. A third used an import-independent order argument and passed both the complete build and the composite contract checker. Retaining all three records explains the choice of the accepted proof.

The final release, \codex{\clmb Double_Tank_Hybrid_UTP_Reconstruction_R10}, kept the earliest nine-theory architecture. The alphabet protected the state and trace vocabulary; the plant theory stated the physical invariant, vector field, continuous evolution, frames, and barrier results; sampling and controller theories established timing, frame, and guard properties; the requirements theory retained the four source requirements and explicit duration semantics; the closed-loop theory composed the operational relations and proved invariant preservation; the conformance theory established tolerance composition and transfer; the monitor theory retained executable observations; and the entry theory exported the public results.

The R9-to-R10 formal delta was confined to \codex{\clmb Double_Tank_Plant.thy}, \codex{\clmb Double_Tank_Requirements.thy}, \codex{\clmb Double_Tank_Closed_Loop.thy}, and \codex{\clmb Double_Tank_Conformance.thy}. All other formal files were protected. The accepted result therefore did not depend on revisions to the alphabet, controller, monitor, imports, or public structure; its changes were limited to the files named in the wider reconstruction contract.

\section{Results and assurance boundary}

The final reconstruction was accepted on 31 July 2026 under Isabelle2025-2 and Poly/ML 5.9.2, containing nine rescue theories and ten scoped reconstruction obligations, all discharged. The complete session and contract-conformance check both passed, and the release contains no \codex{\clmb oops}, \codex{\clmb sorry}, axiomatisations, or oracles. Every accepted delta conformed to its contract, and any change to a theorem statement, semantic definition, architecture, or public assurance claim required the developer to approve a wider contract. This still falls short of complete operational end-to-end verification: trace adequacy remains the separately named obligation DT-R011.

A co-author replayed the complete nine-theory R10 session, Action 34 verifier, and DT-R010 workflow, reproducing the ten discharged obligations, four-file R9--R10 delta, closure markers, admission checks, ten DT-R010 self-tests, and six static Action 8 controls. The co-author also re-examined the 16-declaration audit and its principal end-to-end drift finding. This was an internal secondary replay, not an external replication.

\begin{table}[!h]
  \centering
  \caption{Final reconstruction status.}
  \label{tab:status}
  \small
  \begin{tabularx}{\textwidth}{@{}X>{\raggedleft\arraybackslash}p{0.23\textwidth}@{}}
    \toprule
    Measure & Result \\
    \midrule
    Rescue theories & 9 \\
    Scoped reconstruction obligations & 10 \\
    Discharged scoped obligations & 10 \\
    \codex{\clmb oops} & 0 \\
    \codex{\clmb sorry} & 0 \\
    Axiomatisations & 0 \\
    Oracles & 0 \\
    Complete Isabelle session & Accepted \\
    Contract-conformance verdict & Accepted \\
    \bottomrule
  \end{tabularx}
\end{table}

\noindent R10 does not close every semantic question. Its theorem \codex{\clmb closed_loop_meets_requirements} assumes the predicate \codex{\clmb is_closed_loop_trace} for parameters $\clz p$ and trace $\clz tr$, but not the four requirement predicates themselves. From the explicit clauses of that quantitative trace contract, including strict-safety and uniform-tracking clauses and the adopted duration semantics, R10 derives no-overflow safety, bounded settling, stable tracking, and availability. The open question sits one layer earlier: the release does not yet show that every coherent trace generated by repeated plant, sampling, and controller execution satisfies every clause of \codex{\clmb is_closed_loop_trace}. R10 thus proves the lower arrow in Fig.~\ref{fig:operational-assurance}, while DT-R011 must prove the upper one. ModelPlex addresses an analogous model-to-execution problem through verified runtime validation~\cite{MitschP2014}, albeit in a different logic and toolchain. The next task is therefore the separately named obligation \textbf{DT-R011 Operational Trace Adequacy}, whose intended form is:

\begin{lstlisting}[language=Isabelle,backgroundcolor=\color{gray!10}]
theorem closed_loop_execution_is_valid_trace:
  assumes "admissible_parameters p"
      and "operational_closed_loop_trace p tr"
  shows "is_closed_loop_trace p tr"
\end{lstlisting}

\noindent DT-R011 must begin from a named successor baseline; the accepted R10 release remains immutable. Leaving this boundary open is part of the rescue discipline. An unresolved semantic bridge has been given a stable identifier and an explicit statement, rather than being folded into an assumption or hidden behind an overbroad theorem name.

\section{Discussion}

An Isabelle build certifies the theorem stated in the candidate repository, not preservation of the developer's intended theory frame. CAPRI records that authority boundary and checks candidate deltas against it. A passing contract verdict still leaves the adequacy of an authorised new physical or semantic definition to substantive review and, where necessary, a successor contract. Visible semantic seams are useful only when their later interpretations and downstream effects remain traceable. DT-R011 names and exemplifies this discipline by exposing the missing operational bridge instead of hiding it inside the final theorem's assumptions.

The failed candidates are useful evidence because their diagnostics expose recurring mismatches between informal reasoning and formal obligations, while their deltas reveal when a proposal attempts to change the problem rather than solve it. The project also retains two baselines: the 16-theory repository preserves useful institution and duration machinery and the history of drift, while the nine-theory reconstruction preserves the original task boundary.

The study has important limitations. It reports one hybrid-system development and one reconstruction sequence, without a controlled comparison of repair strategies. Developer judgement influenced prompts, authorised semantic changes, and the retrospective drift classification. The secondary replay was internal; external replication remains future work. The contract checker is trusted tooling and has not itself been verified in Isabelle. The case is also small relative to mature hybrid-verification developments~\cite{QueselMLAP2016,MuniveS2019}; the conclusions are methodological rather than performance claims about large-scale repair.

\section{Conclusion}

The double-tank case exposes a gap between \emph{proof validity} and \emph{task validity}. Isabelle can correctly accept a theory after an iterative repair process weakens a theorem, changes a definition, or strengthens assumptions. The v0.23 audit found material changes in 16 of 100 original declarations; most seriously, the principal assurance theorem derived safety while assuming settling, stability, and availability. The theorem remained valid but no longer discharged the original end-to-end obligation.

The reconstruction shows how to recover from such drift without discarding useful proof knowledge. The original nine-theory frame remains authoritative, while retaining the larger development serves as evidence and a source of lemmas and semantic machinery. Ten reconstruction obligations were discharged under explicit contracts. R10 passes the complete Isabelle build and contract check and contains no \codex{\clmb oops}, \codex{\clmb sorry}, axiomatisations, or oracles. It derives the four requirements from an explicit quantitative trace contract rather than assuming them separately.

A semantic gap remains: R10 does not prove that every trace generated by the plant, sampler, and controller satisfies \codex{\clmb is_closed_loop_trace}. Closing this gap is the explicit successor obligation \textbf{DT-R011, Operational Trace Adequacy}. The broader lesson is that LLM-assisted formal development needs two complementary controls: the prover checks logical validity, while an independent mechanism checks that repair stays within the authorised theory boundary. Semantic adequacy remains a separate engineering judgement.

\section*{AI Usage}

We designed our experiments to explore contract-aware proof repair. We used OpenAI ChatGPT 5.0 to propose Isabelle proof scripts and supporting lemmas in response to prover diagnostics. We did not retain stable model identifiers for every historical ChatGPT interaction. Isabelle2025-2 checked all proof candidates; repository changes in the reconstruction were checked against CAPRI contracts~\cite{WoodcockLSW2026}. We used CAPRI, which invokes OpenAI's Responses API to generate candidate Isabelle proof repairs, using GPT-5.2 (\texttt{\clmb gpt-5.2-2025-12-11}) with high reasoning effort; each repair request is made independently, without shared conversational state. Grammarly, an AI-powered writing assistant, was used to improve grammar, spelling, punctuation, and clarity. The authors reviewed all generated and revised material and take full responsibility for the paper.

\section*{Artefact Availability}

The complete reproducibility artefact is archived at Zenodo, DOI: \url{https://doi.org/10.5281/zenodo.21981425}. The archive contains the source baselines, R10 reconstruction, CAPRI contracts and reports, correspondence and traceability evidence, control suites, replay scripts, manifests, and DT-R011 material. The evaluated release is \codex{\clmb Double_Tank_Hybrid_UTP_Reconstruction_R10}, checked with Isabelle2025-2 and Poly/ML 5.9.2. The recommended replay route is the supplied Action 35 closure package: from \codex{\clmb 03_r10_reconstruction}, verify and extract \codex{\clmb CAPRI_Double_Tank_Action_35_R10_Closure.zip}, then run \codex{\clmb ./verify-action35.sh} and \codex{\clmb ./run-r10-complete.sh}. Alternatively, extract the standalone R10 release and run \texttt{\clmb isabelle build -v -D . -o document=false} from that Isabelle session directory. The Action 36 final-release archive has SHA-256
\begin{lstlisting}[backgroundcolor=\color{gray!10}]
831bc299aa714f17bb304a1c7dd5f37900b5dd11ee03eec3f32bf9cbd850b370
\end{lstlisting}

\bibliographystyle{eptcs}
\bibliography{references}

@inproceedings{Andronick2019,
  author       = {June Andronick},
  editor       = {John Harrison and
                  John O'Leary and
                  Andrew Tolmach},
  title        = {{A} Million Lines of Proof About a Moving Target (Invited Talk)},
  booktitle    = {10th International Conference on Interactive Theorem Proving, {ITP}
                  2019, Portland, OR, USA, September 9-12, 2019},
  series       = {LIPIcs},
  volume       = {141},
  pages        = {1:1--1:1},
  publisher    = {Schloss Dagstuhl - Leibniz-Zentrum f{\"{u}}r Informatik},
  year         = {2019},
  url          = {https://doi.org/10.4230/LIPIcs.ITP.2019.1},
  doi          = {10.4230/LIPICS.ITP.2019.1},
  bibsource    = {dblp computer science bibliography, https://dblp.org}
}

@inproceedings{BohmeN2010,
  author       = {Sascha B{\"{o}}hme and
                  Tobias Nipkow},
  editor       = {J{\"{u}}rgen Giesl and
                  Reiner H{\"{a}}hnle},
  title        = {{Sledgehammer}: {Judgement} {Day}},
  booktitle    = {Automated Reasoning, 5th International Joint Conference, {IJCAR} 2010,
                  Edinburgh, UK, July 16-19, 2010. Proceedings},
  series       = {Lecture Notes in Computer Science},
  volume       = {6173},
  pages        = {107--121},
  publisher    = {Springer},
  year         = {2010},
  url          = {https://doi.org/10.1007/978-3-642-14203-1\_9},
  doi          = {10.1007/978-3-642-14203-1\_9},
  bibsource    = {dblp computer science bibliography, https://dblp.org}
}

@book{ChaochenH2004,
  author       = {Zhou Chaochen and
                  Michael R. Hansen},
  title        = {{Duration} {Calculus} - {A} Formal Approach to Real-Time Systems},
  series       = {Monographs in Theoretical Computer Science. An {EATCS} Series},
  publisher    = {Springer},
  year         = {2004},
  url          = {https://doi.org/10.1007/978-3-662-06784-0},
  doi          = {10.1007/978-3-662-06784-0},
  isbn         = {978-3-642-07404-2},
  bibsource    = {dblp computer science bibliography, https://dblp.org}
}

@article{DeshmukhMP2017,
  author       = {Jyotirmoy V. Deshmukh and
                  Rupak Majumdar and
                  Vinayak S. Prabhu},
  title        = {{Quantifying} conformance using the {Skorokhod} metric},
  journal      = {Formal Methods Syst. Des.},
  volume       = {50},
  number       = {2-3},
  pages        = {168--206},
  year         = {2017},
  url          = {https://doi.org/10.1007/s10703-016-0261-8},
  doi          = {10.1007/S10703-016-0261-8},
  bibsource    = {dblp computer science bibliography, https://dblp.org}
}

@inproceedings{FirstRRB2023,
  author       = {Emily First and
                  Markus N. Rabe and
                  Talia Ringer and
                  Yuriy Brun},
  editor       = {Satish Chandra and
                  Kelly Blincoe and
                  Paolo Tonella},
  title        = {{Baldur}: {Whole}-Proof Generation and Repair with Large Language Models},
  booktitle    = {Proceedings of the 31st {ACM} Joint European Software Engineering
                  Conference and Symposium on the Foundations of Software Engineering,
                  {ESEC/FSE} 2023, San Francisco, CA, USA, December 3-9, 2023},
  pages        = {1229--1241},
  publisher    = {{ACM}},
  year         = {2023},
  url          = {https://doi.org/10.1145/3611643.3616243},
  doi          = {10.1145/3611643.3616243},
  bibsource    = {dblp computer science bibliography, https://dblp.org}
}

@inproceedings{FosterZW2014,
  author       = {Simon Foster and
                  Frank Zeyda and
                  Jim Woodcock},
  editor       = {David A. Naumann},
  title        = {{Isabelle}/{UTP}: {A} Mechanised Theory Engineering Framework},
  booktitle    = {Unifying Theories of Programming - 5th International Symposium, {UTP}
                  2014, Singapore, May 13, 2014, Revised Selected Papers},
  series       = {Lecture Notes in Computer Science},
  volume       = {8963},
  pages        = {21--41},
  publisher    = {Springer},
  year         = {2014},
  url          = {https://doi.org/10.1007/978-3-319-14806-9\_2},
  doi          = {10.1007/978-3-319-14806-9\_2},
  bibsource    = {dblp computer science bibliography, https://dblp.org}
}

@article{GoguenB1992,
  author       = {Joseph A. Goguen and
                  Rod M. Burstall},
  title        = {{Institutions}: {Abstract} Model Theory for Specification and Programming},
  journal      = {J. {ACM}},
  volume       = {39},
  number       = {1},
  pages        = {95--146},
  year         = {1992},
  url          = {https://doi.org/10.1145/147508.147524},
  doi          = {10.1145/147508.147524},
  bibsource    = {dblp computer science bibliography, https://dblp.org}
}

@inproceedings{JiangLTCOMWJ2022,
  author       = {Albert Qiaochu Jiang and
                  Wenda Li and
                  Szymon Tworkowski and
                  Konrad Czechowski and
                  Tomasz Odrzyg{\'{o}}zdz and
                  Piotr Milos and
                  Yuhuai Wu and
                  Mateja Jamnik},
  editor       = {Sanmi Koyejo and
                  S. Mohamed and
                  A. Agarwal and
                  Danielle Belgrave and
                  K. Cho and
                  A. Oh},
  title        = {{Thor}: {Wielding} Hammers to Integrate Language Models and Automated
                  Theorem Provers},
  booktitle    = {Advances in Neural Information Processing Systems 35: Annual Conference
                  on Neural Information Processing Systems 2022, NeurIPS 2022, New Orleans,
                  LA, USA, November 28 - December 9, 2022},
  year         = {2022},
  url          = {http://papers.nips.cc/paper\_files/paper/2022/hash/377c25312668e48f2e531e2f2c422483-Abstract-Conference.html},
  bibsource    = {dblp computer science bibliography, https://dblp.org}
}

@article{LeuckerS2009,
  author       = {Martin Leucker and
                  Christian Schallhart},
  title        = {{A} brief account of runtime verification},
  journal      = {J. Log. Algebraic Methods Program.},
  volume       = {78},
  number       = {5},
  pages        = {293--303},
  year         = {2009},
  url          = {https://doi.org/10.1016/j.jlap.2008.08.004},
  doi          = {10.1016/J.JLAP.2008.08.004},
  bibsource    = {dblp computer science bibliography, https://dblp.org}
}

@inproceedings{MitschP2014,
  author       = {Stefan Mitsch and
                  Andr{\'{e}} Platzer},
  editor       = {Borzoo Bonakdarpour and
                  Scott A. Smolka},
  title        = {{ModelPlex}: {Verified} Runtime Validation of Verified Cyber-Physical
                  System Models},
  booktitle    = {Runtime Verification - 5th International Conference, {RV} 2014, Toronto,
                  ON, Canada, September 22-25, 2014. Proceedings},
  series       = {Lecture Notes in Computer Science},
  volume       = {8734},
  pages        = {199--214},
  publisher    = {Springer},
  year         = {2014},
  url          = {https://doi.org/10.1007/978-3-319-11164-3\_17},
  doi          = {10.1007/978-3-319-11164-3\_17},
  bibsource    = {dblp computer science bibliography, https://dblp.org}
}

@article{MuniveS2019,
  author       = {Jonathan Juli{\'{a}}n Huerta y Munive and
                  Georg Struth},
  title        = {{Predicate} Transformer Semantics for Hybrid Systems: {Verification} Components
                  for {Isabelle}/{HOL}},
  journal      = {CoRR},
  volume       = {abs/1909.05618},
  year         = {2019},
  url          = {http://arxiv.org/abs/1909.05618},
  eprinttype   = {arXiv},
  bibsource    = {dblp computer science bibliography, https://dblp.org}
}

@article{PaulsonNW2019,
  author       = {Lawrence C. Paulson and
                  Tobias Nipkow and
                  Makarius Wenzel},
  title        = {{From} {LCF} to {Isabelle}/{HOL}},
  journal      = {Formal Aspects Comput.},
  volume       = {31},
  number       = {6},
  pages        = {675--698},
  year         = {2019},
  url          = {https://doi.org/10.1007/s00165-019-00492-1},
  doi          = {10.1007/S00165-019-00492-1},
  bibsource    = {dblp computer science bibliography, https://dblp.org}
}

@article{PoluS2020,
  author       = {Stanislas Polu and
                  Ilya Sutskever},
  title        = {{Generative} Language Modeling for Automated Theorem Proving},
  journal      = {CoRR},
  volume       = {abs/2009.03393},
  year         = {2020},
  url          = {https://arxiv.org/abs/2009.03393},
  eprinttype   = {arXiv},
  bibsource    = {dblp computer science bibliography, https://dblp.org}
}

@article{QueselMLAP2016,
  author       = {Jan{-}David Quesel and
                  Stefan Mitsch and
                  Sarah M. Loos and
                  Nikos Ar{\'{e}}chiga and
                  Andr{\'{e}} Platzer},
  title        = {{How} to model and prove hybrid systems with {KeYmaera}: {A} tutorial on
                  safety},
  journal      = {Int. J. Softw. Tools Technol. Transf.},
  volume       = {18},
  number       = {1},
  pages        = {67--91},
  year         = {2016},
  url          = {https://doi.org/10.1007/s10009-015-0367-0},
  doi          = {10.1007/S10009-015-0367-0},
  bibsource    = {dblp computer science bibliography, https://dblp.org}
}

@inproceedings{RingerPYLG2021,
  author       = {Talia Ringer and
                  RanDair Porter and
                  Nathaniel Yazdani and
                  John Leo and
                  Dan Grossman},
  editor       = {Stephen N. Freund and
                  Eran Yahav},
  title        = {{Proof} repair across type equivalences},
  booktitle    = {{PLDI} '21: 42nd {ACM} {SIGPLAN} International Conference on Programming
                  Language Design and Implementation, Virtual Event, Canada, June 20-25,
                  2021},
  pages        = {112--127},
  publisher    = {{ACM}},
  year         = {2021},
  url          = {https://doi.org/10.1145/3453483.3454033},
  doi          = {10.1145/3453483.3454033},
  bibsource    = {dblp computer science bibliography, https://dblp.org}
}

@inproceedings{ThaeH2001,
  author       = {Hong Ki Thae and
                  Dang Van Hung},
  title        = {{A} Case Study on Formal Design of Hybrid Control Systems},
  booktitle    = {25th International Computer Software and Applications Conference {(COMPSAC}
                  2001), Invigorating Software Development, 8-12 October 2001, Chicago,
                  IL, {USA}},
  pages        = {423--428},
  publisher    = {{IEEE} Computer Society},
  year         = {2001},
  url          = {https://doi.org/10.1109/CMPSAC.2001.960648},
  doi          = {10.1109/CMPSAC.2001.960648},
  bibsource    = {dblp computer science bibliography, https://dblp.org}
}

@inproceedings{YangSGCSYGPA2023,
  author       = {Kaiyu Yang and
                  Aidan M. Swope and
                  Alex Gu and
                  Rahul Chalamala and
                  Peiyang Song and
                  Shixing Yu and
                  Saad Godil and
                  Ryan J. Prenger and
                  Animashree Anandkumar},
  editor       = {Alice Oh and
                  Tristan Naumann and
                  Amir Globerson and
                  Kate Saenko and
                  Moritz Hardt and
                  Sergey Levine},
  title        = {{LeanDojo}: {Theorem} Proving with Retrieval-Augmented Language Models},
  booktitle    = {Advances in Neural Information Processing Systems 36: Annual Conference
                  on Neural Information Processing Systems 2023, NeurIPS 2023, New Orleans,
                  LA, USA, December 10 - 16, 2023},
  year         = {2023},
  url          = {http://papers.nips.cc/paper\_files/paper/2023/hash/4441469427094f8873d0fecb0c4e1cee-Abstract-Datasets\_and\_Benchmarks.html},
  bibsource    = {dblp computer science bibliography, https://dblp.org}
}

@inproceedings{foster2019hybrid,
  author       = {Simon Foster},
  editor       = {Pedro Ribeiro and
                  Augusto Sampaio},
  title        = {{Hybrid} Relations in {Isabelle}/{UTP}},
  booktitle    = {Unifying Theories of Programming - 7th International Symposium, {UTP}
                  2019, Dedicated to Tony Hoare on the Occasion of His 85th Birthday,
                  Porto, Portugal, October 8, 2019, Proceedings},
  series       = {Lecture Notes in Computer Science},
  volume       = {11885},
  pages        = {130--153},
  publisher    = {Springer},
  year         = {2019},
  url          = {https://doi.org/10.1007/978-3-030-31038-7\_7},
  doi          = {10.1007/978-3-030-31038-7\_7},
  bibsource    = {dblp computer science bibliography, https://dblp.org}
}

@inproceedings{GotelF1994,
  author    = {Orlena C. Z. Gotel and Anthony C. W. Finkelstein},
  title     = {An Analysis of the Requirements Traceability Problem},
  booktitle = {Proceedings of the First IEEE International Conference on Requirements Engineering},
  pages     = {94--101},
  publisher = {IEEE Computer Society},
  year      = {1994},
  doi       = {10.1109/ICRE.1994.292398},
  url       = {https://doi.org/10.1109/ICRE.1994.292398}
}

@inproceedings{NuseibehE2000,
  author    = {Bashar Nuseibeh and Steve M. Easterbrook},
  title     = {Requirements Engineering: A Roadmap},
  booktitle = {Proceedings of the Conference on the Future of Software Engineering},
  pages     = {35--46},
  publisher = {ACM},
  year      = {2000},
  doi       = {10.1145/336512.336523},
  url       = {https://doi.org/10.1145/336512.336523}
}

@article{RingerPSGT2019,
  author  = {Talia Ringer and Karl Palmskog and Ilya Sergey and Milos Gligoric and Zachary Tatlock},
  title   = {{QED} at Large: A Survey of Engineering of Formally Verified Software},
  journal = {Foundations and Trends in Programming Languages},
  volume  = {5},
  number  = {2--3},
  pages   = {102--281},
  year    = {2019},
  doi     = {10.1561/2500000045},
  url     = {https://doi.org/10.1561/2500000045}
}

@techreport{WoodcockLSW2026,
  author      = {Jim Woodcock and
                 Gabriel Leite and
                 Augusto Sampaio and
                 Ran Wei},
  title       = {{CAPRI}: {Contract}-Aware Proof Repair for {Isabelle}},
  institution = {University of York},
  year        = {2026},
  type        = {Technical Report},
  address     = {UK},
}

\end{document}